\pdfoutput=1 

\documentclass[12pt]{article}
\usepackage{amsmath,graphicx,hhline,siunitx}
\usepackage{geometry} 
\date{}

\begin{document}

\author{Francesco Grilli$^1$, Antonio Morandi$^2$, Federica De Silvestri$^2$,\\ Roberto Brambilla$^3$\\
$^1${\small Karlsruhe Institute of Technology, Karlsruhe, Germany} \\
$^2${\small University of Bologna, Bologna, Italy} \\
$^3${\small Ricerca sul Sistema Energetico, Milan, Italy}
}
\title{Dynamic modeling of levitation of a superconducting bulk by coupled $H$-magnetic field and Arbitrary Lagrangian-Eulerian formulations}

\maketitle

{
}

\begin{abstract}
Intrinsically stable magnetic levitation between superconductors and permanent magnets can be exploited in a variety of applications of great technical interest in the field of transportation (rail transportation), energy (flywheels) and industry. In this contribution, we present a new model for the calculation of levitation forces between superconducting bulks and permanent magnet, based on the $H$-formulation of Maxwell's equations coupled with an Arbitrary Lagrangian-Eulerian formulation. The model uses a moving mesh that adapts at each time step based on the time-change of the distance between a superconductor bulk and a permanent magnet. The model is validated against a fixed mesh model (recently in turn validated against experiments)  that uses an analytical approach for calculating the magnetic field generated by the moving permanent magnet. Then, it is used to analyze the magnetic field dynamics both in field-cooled and zero-field-cooled conditions and successively used to test different configurations of permanent magnets and to compare them in terms of levitation forces. 
The easiness of implementation of this model and its flexibility in handling  different geometries, material properties, and application scenarios make the model an attractive tool for the analysis and optimization of magnetic levitation-based applications.
\end{abstract}




\section{Introduction}

Magnetic levitation between permanent magnets and superconductor bulks can be exploited in variety of applications of great practical interest, including rail transportation~\cite{Werfel:SST12,Sotelo:TAS15} and flywheel energy storage~\cite{Werfel:SST12, Strasik:SST09, Miyazaki:Cryo16}. Numerical modeling is the main tool available for understanding the electrodynamics of the observed levitation phenomena and for predicting and optimizing the performance practical levitation apparatus. In the past years a great research effort has been devoted to the development of numerical formulations able to simulate the behavior of levitation systems~\cite{Zhou:PhysC06, Perini:TAS09, Morandi:TAS11, Navau:TAS13, Wu:PhysC13, Azzouza:TMAG17, Bernstein:SST17} including models based on the $H$-formulation of Maxwell's equations~\cite{Sass:SST15, Patel:SST17, Sass:SST18}.

The $H$-formulation has become a popular tool for investigating the electrodynamic behavior of superconductors, in particular the AC losses -- see for example~\cite{Nguyen:SST11, Ainslie:PhysC12, Zhang:JAP12, Xia:JAP13, IftekharJaim:JLTP13, Zhao:Cryo17}. 
When used to simulate levitation problems (as in \cite{Sass:SST15, Patel:SST17, Sass:SST18}), the relative movement of the permanent magnet with respect to the superconductor is accounted for by simulating only the superconductor and setting  appropriate time-dependent boundary conditions for the magnetic field generated by the permanent magnet. One disadvantage of this approach is that the boundary conditions have to be calculated analytically, and -- in general -- change with the particular problem under consideration, e.g. 2D axisymmetric, 2D cartesian (infinite length) with or without symmetries, etc.

In the present work, we present a new approach that simulates the whole geometry (superconductor and permanent magnet) and calculates the electromagnetic interaction directly in the $H$-formulation. The relative movement of superconductor and permanent magnet is accounted for by means of a deforming mesh. This approach allows for a great flexibility of the types of problems that can be solved and it provides a ``ready-to-run'' model that can be easily changed according to the different simulation scenarios. This latter point is particularly interesting for users who do not have much resources to invest in the development of an {\it ad hoc} model every time. 

\section{Problem definition and numerical model}
In this work we calculate the levitation force between a cylindrical permanent magnet and coaxial cylindrical high-temperature superconductor (HTS) bulk. Both field-cooled (FC) and zero field-cooled (ZFC) conditions are considered. The geometrical and physical parameters of the problems are listed in table~\ref{tab:param}. 

\begin{table}[t!]
\renewcommand{\arraystretch}{1.1}
\begin{center}
 \caption{Physical and geometrical parameters}\label{tab:param}
\begin{tabular}{lll}
Name & Description & Value\\
\hline
$E_{\rm c}$	&	Critical electrical field	&	\SI{1e-4}{\volt\per\meter}	\\
$n$			&	Power index			&	40					\\
$J_{\rm c}(\SI{75}{\kelvin}, \SI{0}{\tesla})$	&	Critical current density at \SI{75}{\kelvin}, \SI{0}{\tesla}	&	\SI{1.89e8}{\ampere\per\meter\squared} \\
$J_{\rm c}(\SI{75}{\kelvin}, \SI{0.6}{\tesla})$	&	Critical current density at \SI{75}{\kelvin}, \SI{0.6}{\tesla}	&	\SI{1.35e8}{\ampere\per\meter\squared} \\
$\rho_{\rm n}$	&	Normal state resistivity	&	\SI{1e-6}{\ohm\meter} 	\\
$M_0$		&	Magnetization of permanent magnet & \SI{6.6903e5}{\ampere\per\meter} \\
$r_{\rm sc}$	&	Radius of superconductor bulk	&	\SI{12.5}{\milli\meter} \\
$r_{\rm pm}$	&	Radius of permanent magnet	&	\SI{12.5}{\milli\meter} \\
$h_{\rm sc}$	&	Height of superconductor bulk	&	\SI{18}{\milli\meter} \\
$h_{\rm pm}$	&	Height of permanent magnet	&	\SI{15}{\milli\meter} \\
$g_{\rm ZFC}$	&	Initial gap (ZFC)			&	\SI{46.81}{\milli\meter} \\
$g_{\rm FC}$	&	Initial gap (FC)				&	\SI{0.1}{\milli\meter} \\
$d$			&	Excursion (ZFC and FC)		&	\SI{46.71}{\milli\meter} \\
$t_0$		&	Time for one-way run		&	\SI{122.92}{\second}	\\
 \end{tabular}
\end{center}
\end{table}

The  interaction between the superconductor and the permanent magnet is calculated by a finite-element method (FEM) model employing the $H$-formulation of Maxwell's equations for the electromagnetic part and the Arbitrary Lagrangian-Eulerian (ALE) formulation for taking the movement into account.
The model is implemented in the Comsol Multiphysics FEM software package.
\subsection{$H$-formulation} 
The $H$-formulation of Maxwell's equations uses edge elements shape functions for approximating the magnetic field over the computation domain. The state variables of the discretized problem are the components of the magnetic field along the edges of the mesh~\cite{Pecher:EUCAS2003}. Its implementation in the commercial software Comsol Multiphysics~\cite{Brambilla:SST07} has become a popular tool for investigating the electrodynamic behavior of superconductors, in particular the AC losses -- see for example~\cite{Nguyen:SST11, Ainslie:PhysC12, Zhang:JAP12, Xia:JAP13, IftekharJaim:JLTP13, Zhao:Cryo17}. In its original implementation and in most works published in the literature, the equations are implemented in Comsol's  general module for partial differential equations.\footnote{The precise name is PDE, General Form} In recent versions of the software, the $H$-formulation has become directly available as a built-in module, called MFH. In this module, the electromagnetic quantities derived from the magnetic field components (which constitute the state variables), such as the current density and the electric field, are automatically defined. The main advantage, however, is that in the MFH module it is possible to easily define the magnetic properties of a material, by assigning $B-H$ curves or a given magnetization $M_0$. The latter is what is needed to simulate permanent magnets.

The superconductor is modeled as a material with non-linear resistivity given by a power-law relationship between the electric field and the current density~\cite{Rhyner:PhysC93}. In order to avoid a non-physical divergence of the electrical field $E$ to infinity for large value of $J$, the power-law is limited with the normal state resistivity of the superconductor $\rho_{\rm n}$~\cite{Duron:PhysC04,Fabbri:TAS06,Ainslie:SST16}. An analogous relation was introduced in~\cite{Brandt:PRB99a} in order to link the flux creep to the flux flow regime. This feature gives a more realistic representation of the electrical resistivity of a superconductor for over-critical current densities and, by avoiding the possibility of reaching very large values for the resistivity, helps increase the convergence of calculations.

The expression for the resistivity is 
\begin{equation}
\rho=\frac{\rho_{\rm PL}\rho_{\rm n}}{\rho_{\rm PL}+\rho_{\rm n}}
\end{equation}
with
\begin{equation}\label{eq:PL}
\rho_{\rm PL}=\frac{E_{\rm c}}{J_{\rm c}(B)} \left | \frac{J}{J_{\rm c}(B)}\right |^{n-1}
\end{equation}
where $J_{\rm c}(B)$ is the linear interpolation of the data from experiments presented in table~\ref{tab:param}.
The permanent magnet is simulated as a non-conducting material, with permanent magnetization $M_0$. Unless specified otherwise, the magnetization is assumed to be uniform in the whole volume of the permanent magnet.

A preliminary magnetostatic analysis is necessary, in order to calculate the initial field inside the superconducting bulk.
First, a static simulation is performed, where the bulk is assigned the electromagnetic properties of air, thus allowing the magnetic field produced by the permanent magnet penetrate in the bulk's interior. Then, the obtained solution is  used as initial condition for the dynamic simulation, where the bulk is assigned superconducting properties (equation~\eqref{eq:PL}). In case of strict ZFC conditions, i.e. with an infinite initial gap between the superconductor and the permanent, a zero magnetic field is found on the SC bulk at the initial instant. The same occurs in practice with the actual ZFC problem dealt with in this paper, since a negligible field on the superconductor is found at the begin when the initial gap is quite large (\SI{46.81}{\milli\meter}). This means that, in practice, the initial static simulation can be avoided, and one can just simulate the dynamic problem.

In all cases, the instantaneous levitation force is calculated as
\begin{equation}\label{eq:force}
F_{\rm l}=\int\limits_{S} 2 \pi r J_{\phi} B_r {\rm d}S,
\end{equation}
where $J_{\phi}$ is the azimuthal component of the current density, $B_r$ the radial component of the magnetic flux density and $S$ is the bulk's cross-section in the $rz$ plane. The instantaneous power dissipation is calculated as
\begin{equation}\label{eq:power}
P=\int\limits_{S} 2 \pi r J_{\phi} E_{\phi} {\rm d}S,
\end{equation}
where $E_{\phi}$ is the azimuthal component of the electric field.

\subsection{Moving mesh}
The moving mesh is obtained with Comsol's ALE module. Given the axial symmetry of the problem, the problem is simulated in 2D cylindrical coordinates. Figure~\ref{fig:ALE} shows the definition of the different domains (left) and boundaries (right). The superconducting bulk and the permanent magnet correspond to domains 2 and 4, respectively. All the other domains are considered as air. The blue domains move, following the displacement of the bulk. This is realized by assigning different properties to the various boundaries. 
In particular, the boundaries associated with the area of the permanent magnet (7, 8, 10, 18) are assigned a time-dependent displacement in the z direction and null displacement in the radial direction. The same is done for boundaries 21, 26, 19, which delimit the region of air with the same $z$-coordinate as the permanent magnet (domain 9 in figure~\ref{fig:ALE}). The axial length of boundaries 9, 10, 27 and 5, 16, 25, deform with time (shrink or expand) following the movement of the permanent magnet. All the other boundaries in figure~\ref{fig:ALE} remain fixed during the simulation.

\subsection{Effect of gravity}\label{subsec:grav}
In most of the cases presented here, both the axial position $z$ and the velocity $v$ of the PM are assigned quantities.
However, the model can be adapted to the case when $z$ and $v$ are not known a priori but are obtained by solving the equation of motion of the permanent magnet. An example where the permanent magnet is left free to fall toward the superconductor bulk is given in section~\ref{subsec:grav}.

\begin{figure}[t!]
\centering
\includegraphics[width=8 cm]{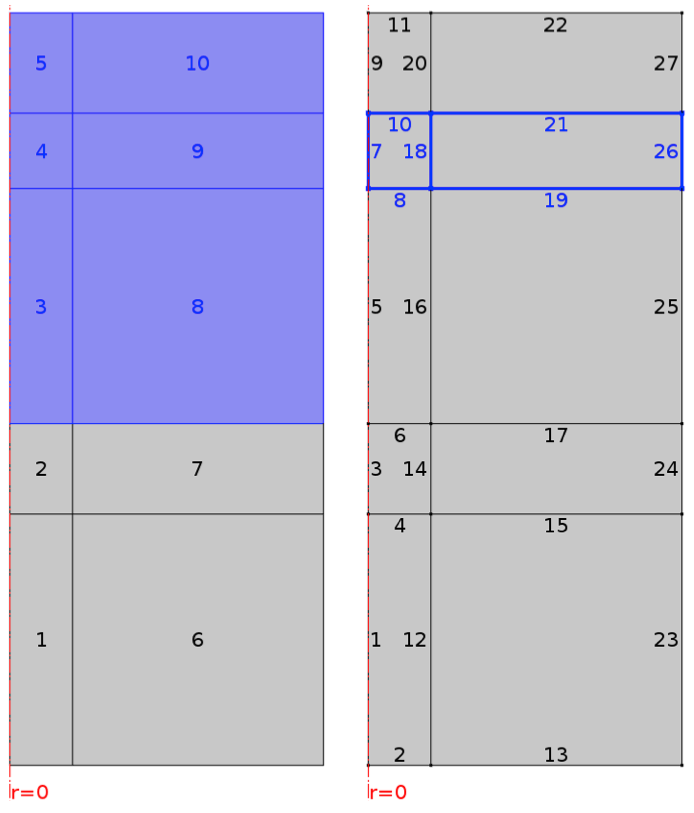}
\caption{\label{fig:ALE}Domains (left) and boundaries (right) of the problem. Given the axial symmetry, the problem is simulated in 2D cylindrical coordinates. The superconducting bulk and the permanent magnet correspond to domains 2 and 4, respectively. The blue domains change with time, following the movement of the blue-colored boundaries.}
\end{figure}

The introduction of two additional state variables of the time domain problem, the position $z$ and velocity $v$ of the permanent magnet, and, consequently, of two additional equations. One is the equation of motion, expressed as 
\begin{equation}
\dot{v}_z=\frac{F_{\rm l}+F_{\rm g}}{m_{\rm PM}}
\end{equation}
where $F_{\rm l}$ and $F_{\rm g}$ are the levitation and gravity forces, respectively, and $m_{\rm PM}$ is the mass of the permanent magnet; the other is the definition of the velocity of the permanent magnet
\begin{equation}
\dot{z}-v=0,
\end{equation}
with initial conditions $z=z_0$ and $v=v_0$. The mesh displacement for the blue boundaries in figure~\ref{fig:ALE}) is in this case simply $z-z_0$.

\section{Results}
\subsection{Model validation}
In order to validate our model, we compared the results with those of a recently proposed model based on the $A-\phi$ formulation of the eddy current problem~\cite{Morandi:TAS18}, which has been in turn validated with comparison with experimental data. This is in essence a 2D axisymmetric finite element model obtained by discretizing the superconductor cylinder into a finite number of loops and by applying the volume integral equation method (VIEM) for obtaining the solving system [11]

\begin{figure}[t!]
\centering
\includegraphics[width=6 cm]{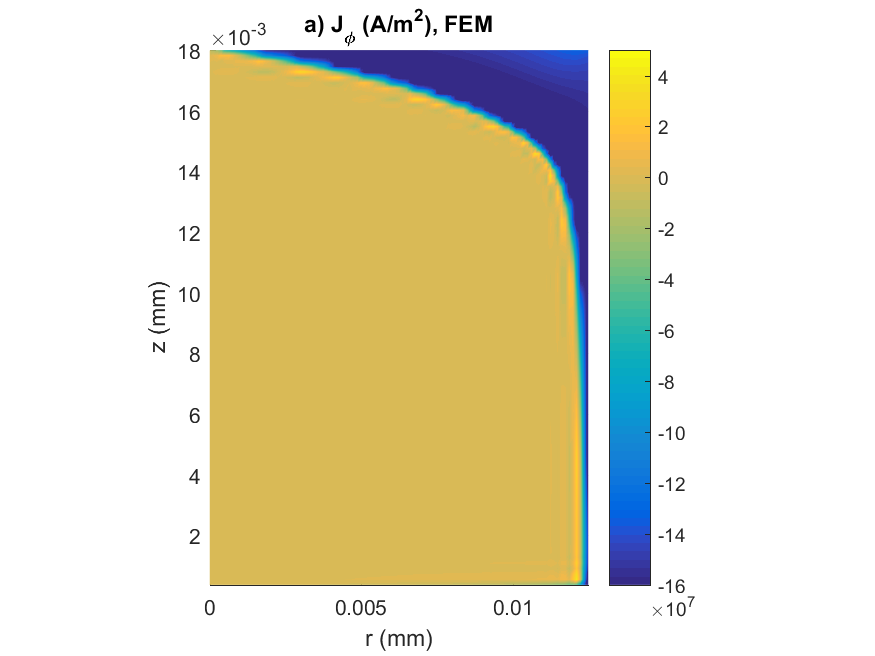}
\includegraphics[width=6 cm]{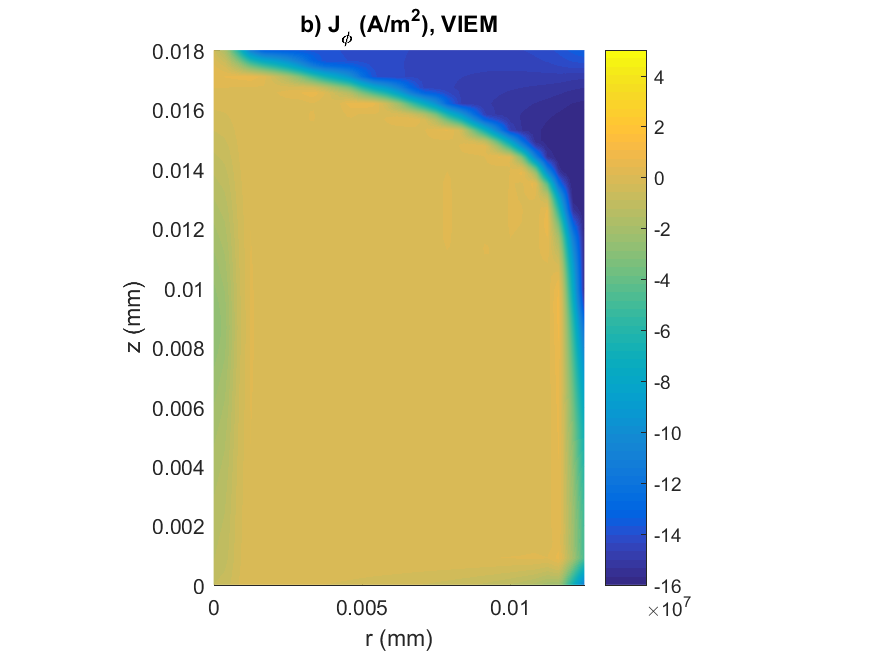}
\includegraphics[width=6 cm]{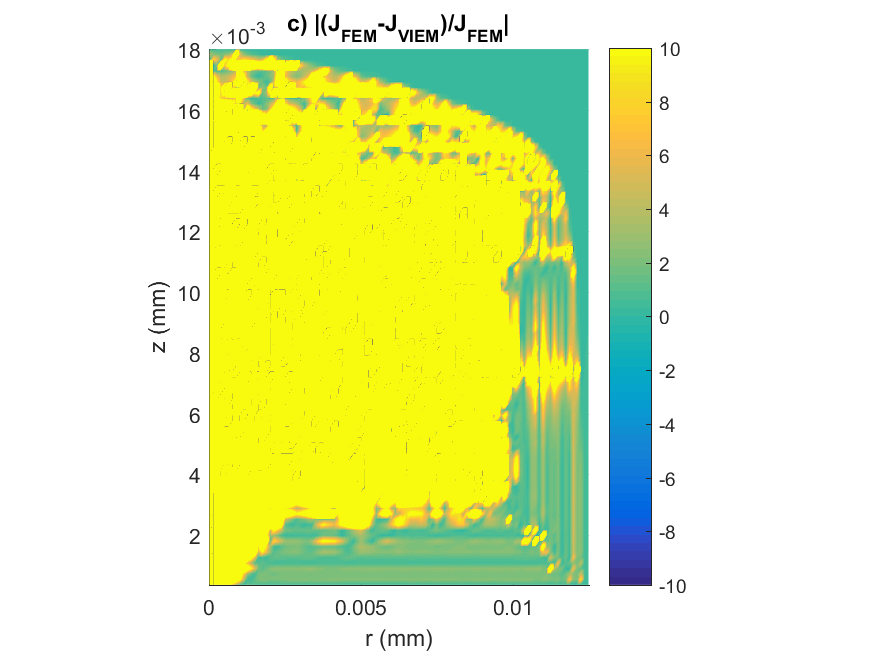}
\includegraphics[width=6 cm]{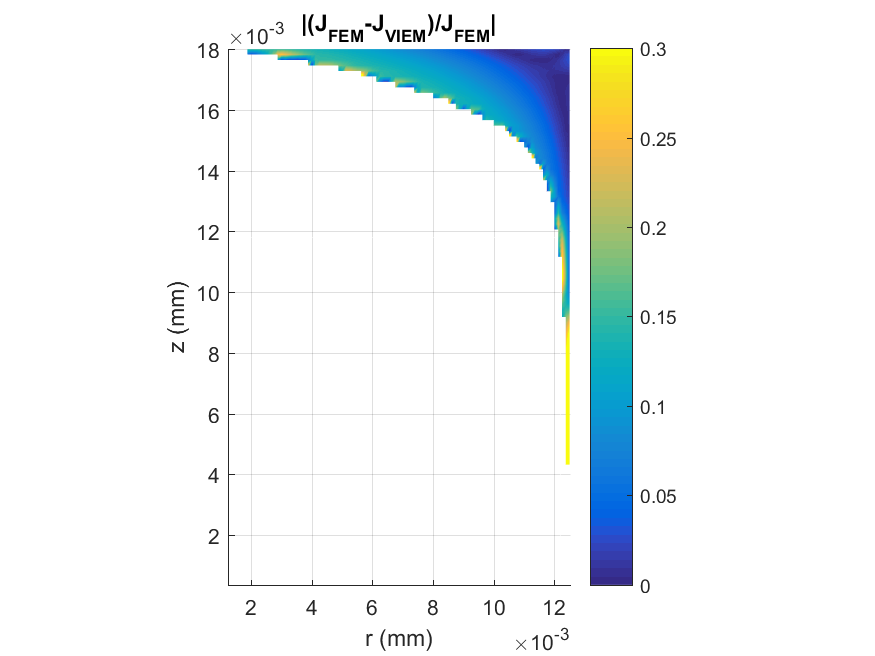}
\caption{\label{fig:J_comparison}Comparison of $J$ distributions in the superconductor computed with the FEM and VIEM models at the end of the first run of the ZFC case, at the instant of minimum distance. In figure figure~\ref{fig:J_comparison}d, the difference is plotted only in the in the regions where there is a substantial current density (higher than \SI{1e8}{\ampere\per\meter\squared}).}
\end{figure}

Figure~\ref{fig:J_comparison} compares the current density distribution in the superconductor computed with the FEM and VIEM models at the end of the first run of the ZFC case, at the instant $t=t_0$, where $t_0$ is the time for a one-way run (table~\ref{tab:param}). The results are very similar, and the differences are quantified in figures~\ref{fig:J_comparison}c-d. As shown in figure~\ref{fig:J_comparison}c, the largest differences occur in the current-free region of the bulk: this is because the current density assumes values very close to zero and the calculated difference is inevitably very large. On the contrary, in the current-carrying region, where the magnetic flux has penetrated, the difference is much smaller: \SI{10.7}{\percent} on average, with higher differences being confined to the thin boundary of the current-carrying region. This is visualized in figure~\ref{fig:J_comparison}d, where the error is calculated only in the in the regions where there is a substantial current density (higher than \SI{1e8}{\ampere\per\meter\squared}).

\begin{figure}[t!]
\centering
\includegraphics[width=8 cm]{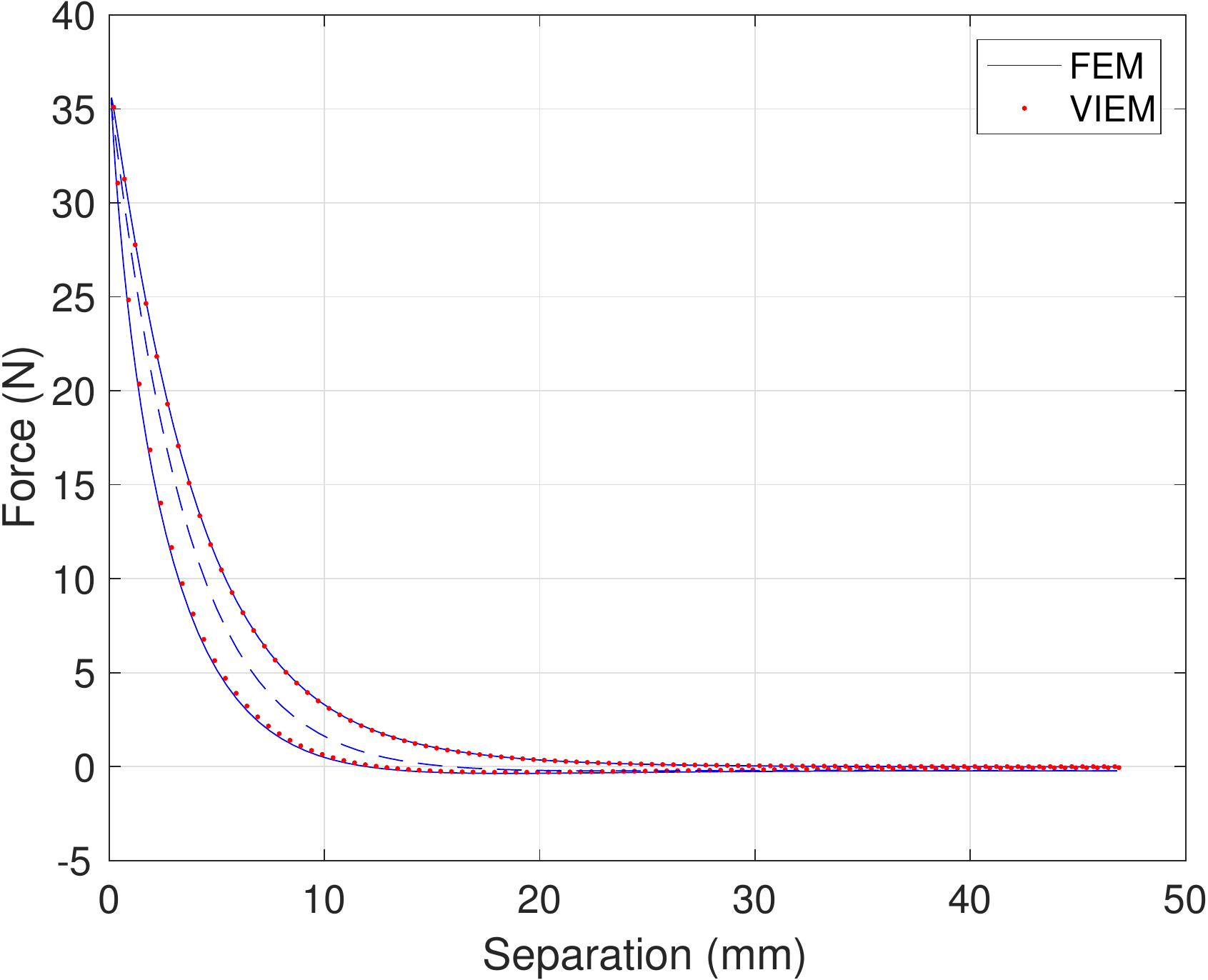}
\caption{\label{fig:force}Levitation force between the superconductor and permanent magnet as a function of the distance between them during the first cycle $(0, 2t_0)$ calculated with the FEM and VIEM models in the zero-field-cooled (ZFC) situation. The dashed line represents the force during the interval $(2t_0, 3t_0)$.}
\end{figure}

Figure~\ref{fig:force} shows the levitation force as a function of the distance between the bulk and the permanent magnet during the first cycle $(0, 2t_0)$. The curves of the forces calculated with the two models overlap. Shown in the figure is also the force calculated in the interval $(2t_0, 3t_0)$. The force is different from that in $(0, t_0)$ because the superconducting bulk is no longer in the virgin state. For $t>3t_0$ the force has a cyclic behavior. The steady state behavior of the system starts after the first magnetization is accomplished by means of the firs run of the permanent magnet. This means that in order to calculate the ``steady-state'' behavior of the system, it is necessary to simulate at least 1.5 cycles.

\begin{figure}[t!]
\centering
\includegraphics[width=8 cm]{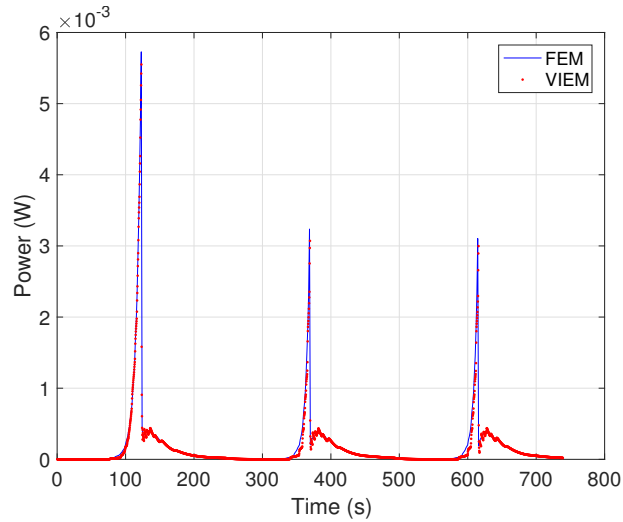}
\caption{\label{fig:power}Power dissipated in the superconducting bulk as a function of time during the three cycles $(0, 6t_0)$ calculated with the FEM and VIEM models in the zero-field-cooled (ZFC) situation.}
\end{figure}

Figure~\ref{fig:power} shows the power dissipation in the superconductor bulk as a function of time over three cycles. Also from this figure can one see that the first cycle is not representative of the cyclic behavior, as a greater power is dissipated in the superconductor bulk during the first cycle, when it is in the virgin sate. 

\begin{figure}[t!]
\centering
\includegraphics[width=8 cm]{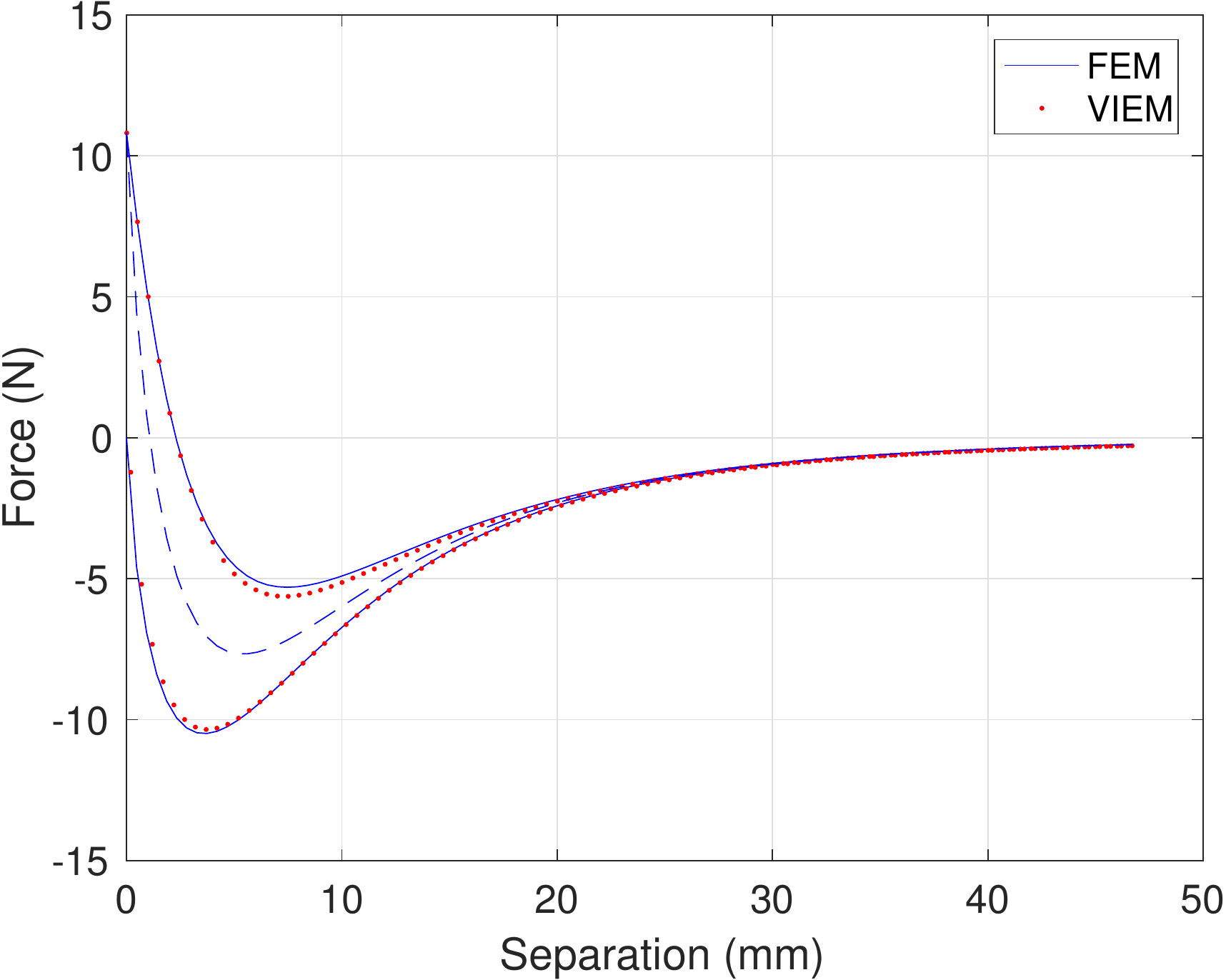}
\caption{\label{fig:FC_force}Levitation force between the superconductor and permanent magnet as a function of the distance between them during the first cycle $(0, 2t_0)$ calculated with the FEM and VIEM models in the field-cooled (FC) situation. The dashed line represents the force during the interval $(2t_0, 3t_0)$.}
\end{figure}

\begin{figure}[t!]
\centering
\includegraphics[width=8 cm]{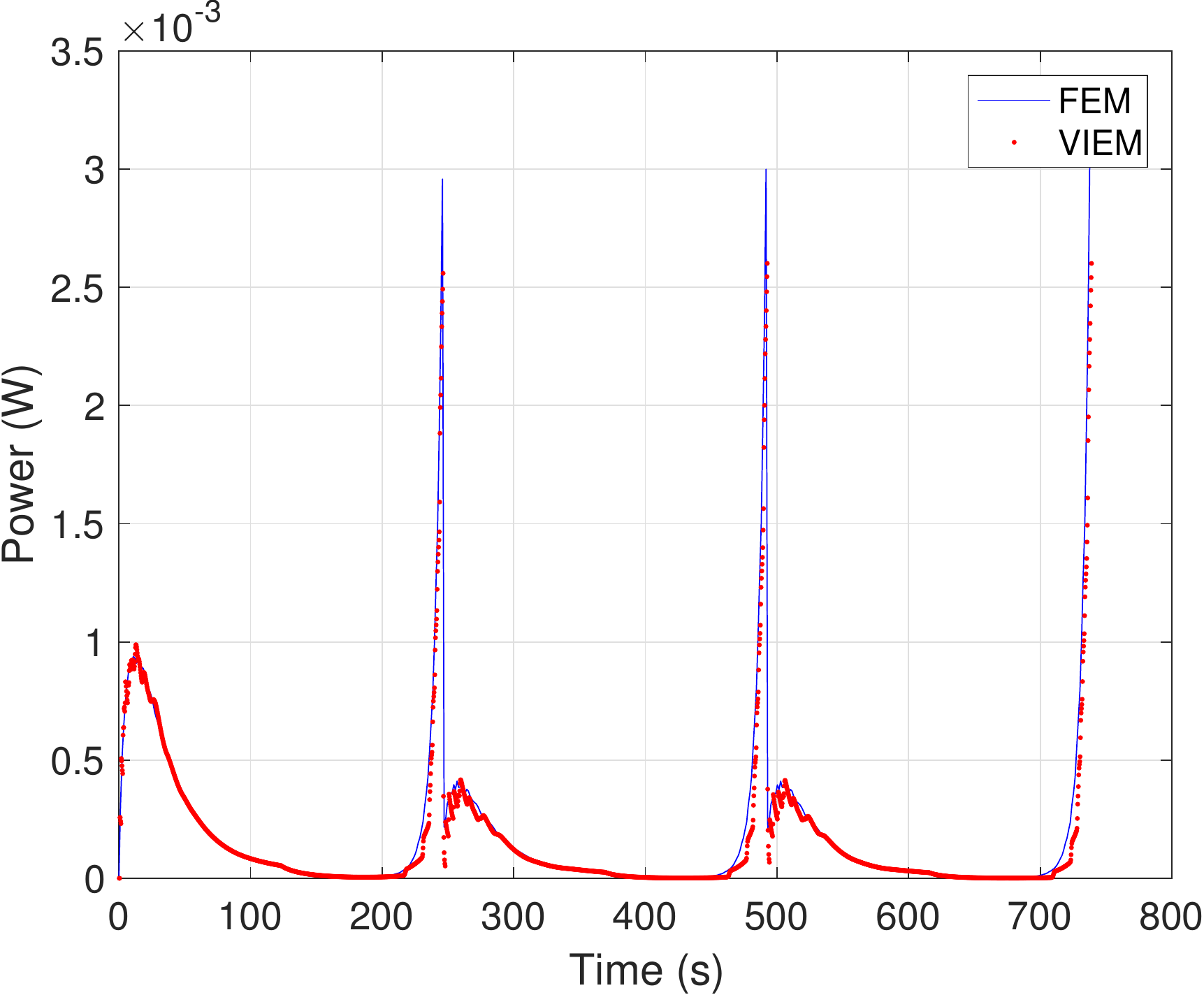}
\caption{\label{fig:FC_power}Power dissipated in the superconducting bulk as a function of time during the three cycles $(0, 6t_0)$ calculated with the FEM and VIEM models in the field-cooled (FC) situation.}
\end{figure}

Figures~\ref{fig:FC_force} and~\ref{fig:FC_power} show the levitation force  and the power dissipation in the superconductor bulk over three cycles for the field-cooling case. A peak of the (negative) levitation force is observed in figure~\ref{fig:FC_force} at a separation of about \SI{4}{\milli\meter}, due to the increase of the total current induced in the bulk and the decrease of the field produced on it as the permanent magnet moves away. A lower mechanical interaction is observed between the bulk and the permanent magnet in the FC case (figure~\ref{fig:FC_force}) compared with the ZFC case (figure~\ref{fig:force}). This means that a lower current is induced in the superconductor in the first case. Consistently, a lower power is dissipated inside it as it can be confirmed by the comparison of figures~\ref{fig:power} and~\ref{fig:FC_power}. 

\subsection{Application examples}
The developed model can be easily used to investigate several cases of practical interest, involving for example different configurations of permanent magnets and soft magnetic materials e.g for the purpose of increasing the levitation force. Another case is the coupling of the  model with the equation of motion of the permanent magnet as discussed in section~\ref{subsec:grav}. Some of these cases are discussed here below, with the purpose of showing the flexibility of application and the efficiency of the developed model.
\subsubsection{Opposite magnetizations}
\begin{figure}[t!]
\centering
\includegraphics[width=8 cm]{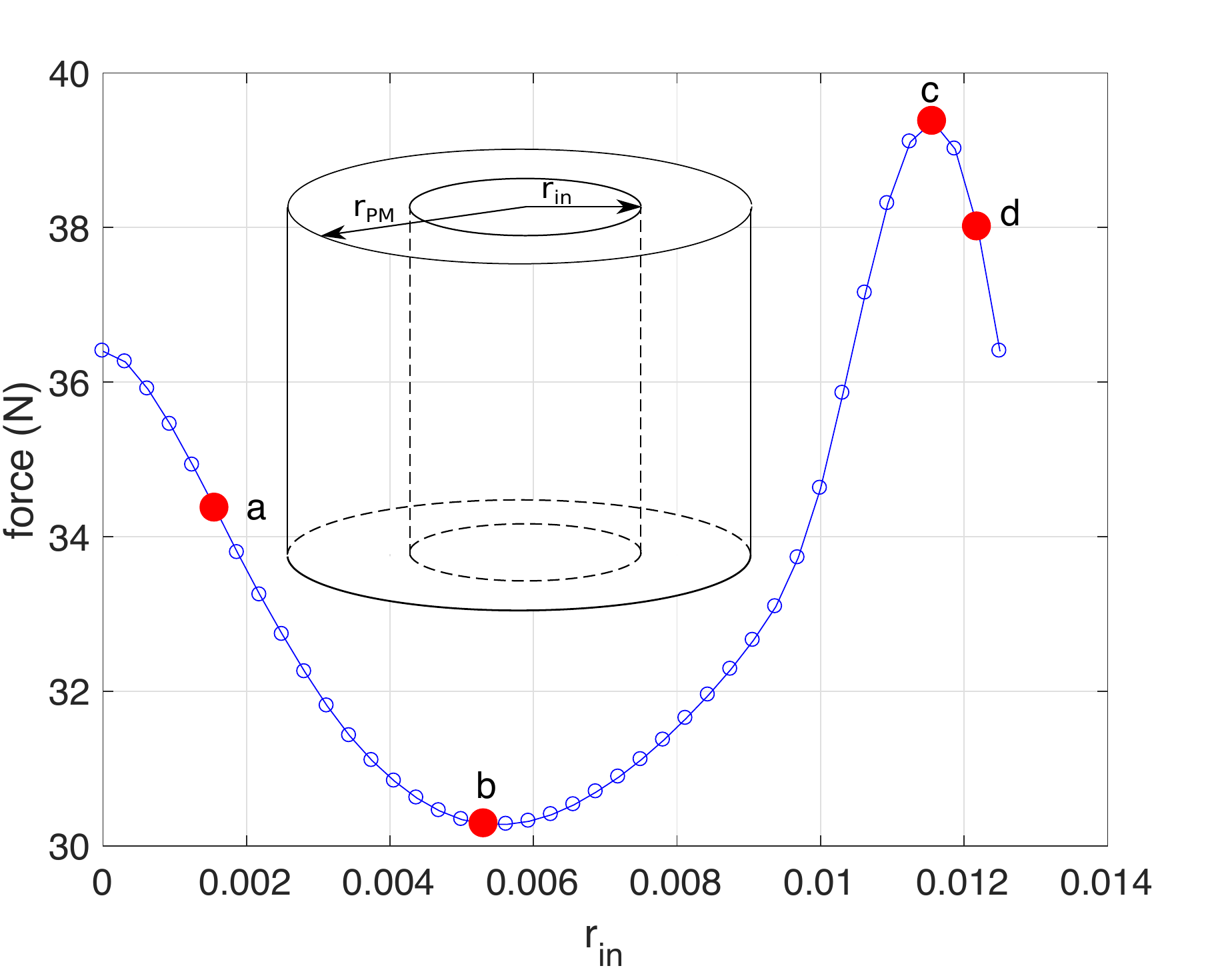}
\caption{\label{fig:magn_up_down}Maximum force obtained with a permanent magnet composed of two concentric cylinders with opposite magnetizations $+M_0$ and $-M_0$ in the $z$ direction. The force shown refers to the minimum distance (\SI{0.1}{\milli\meter}) between  the superconductor and permanent magnet. The parameter $r_{\rm in}$ is the radius of the inner cylinder with magnetization $+M_0$.}
\end{figure}

\begin{figure}[h!]
\centering
\includegraphics[width=6 cm]{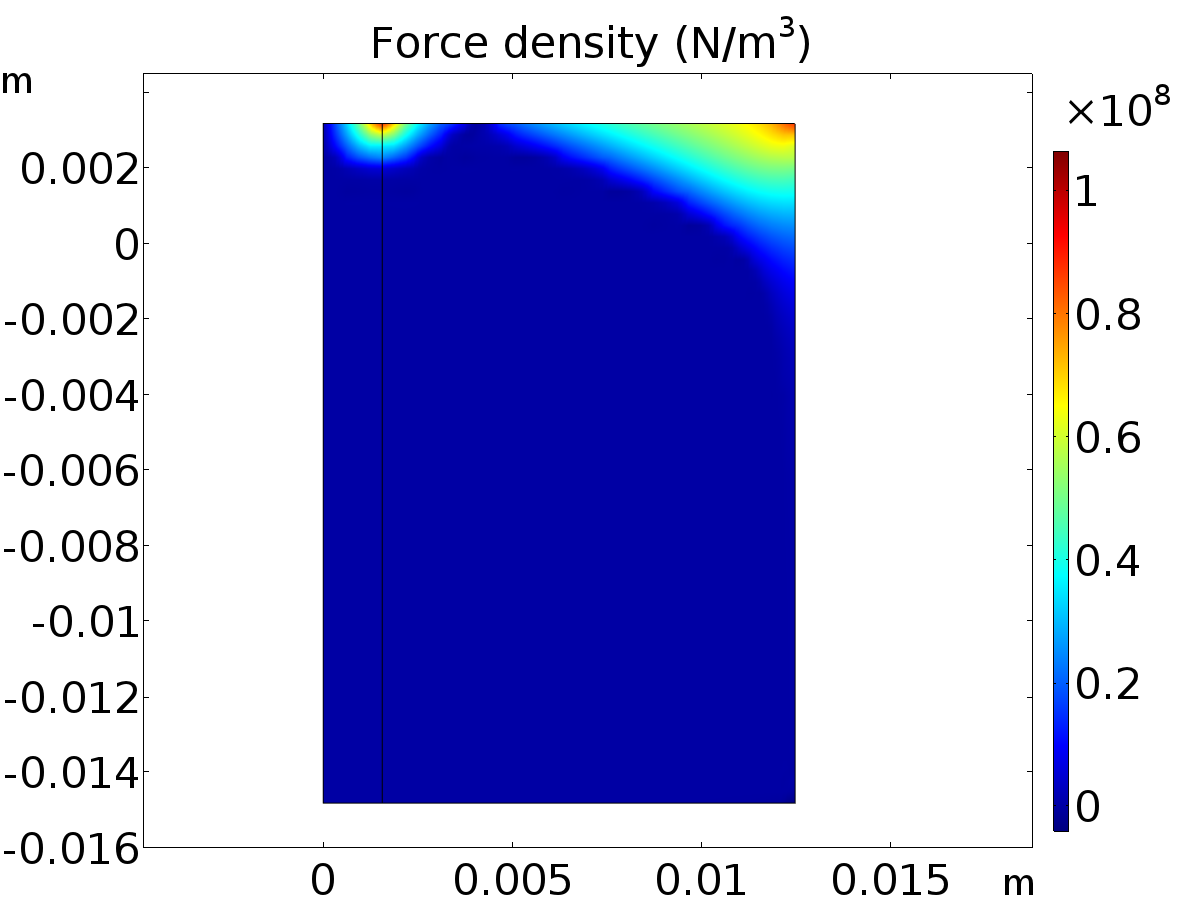}
\includegraphics[width=6 cm]{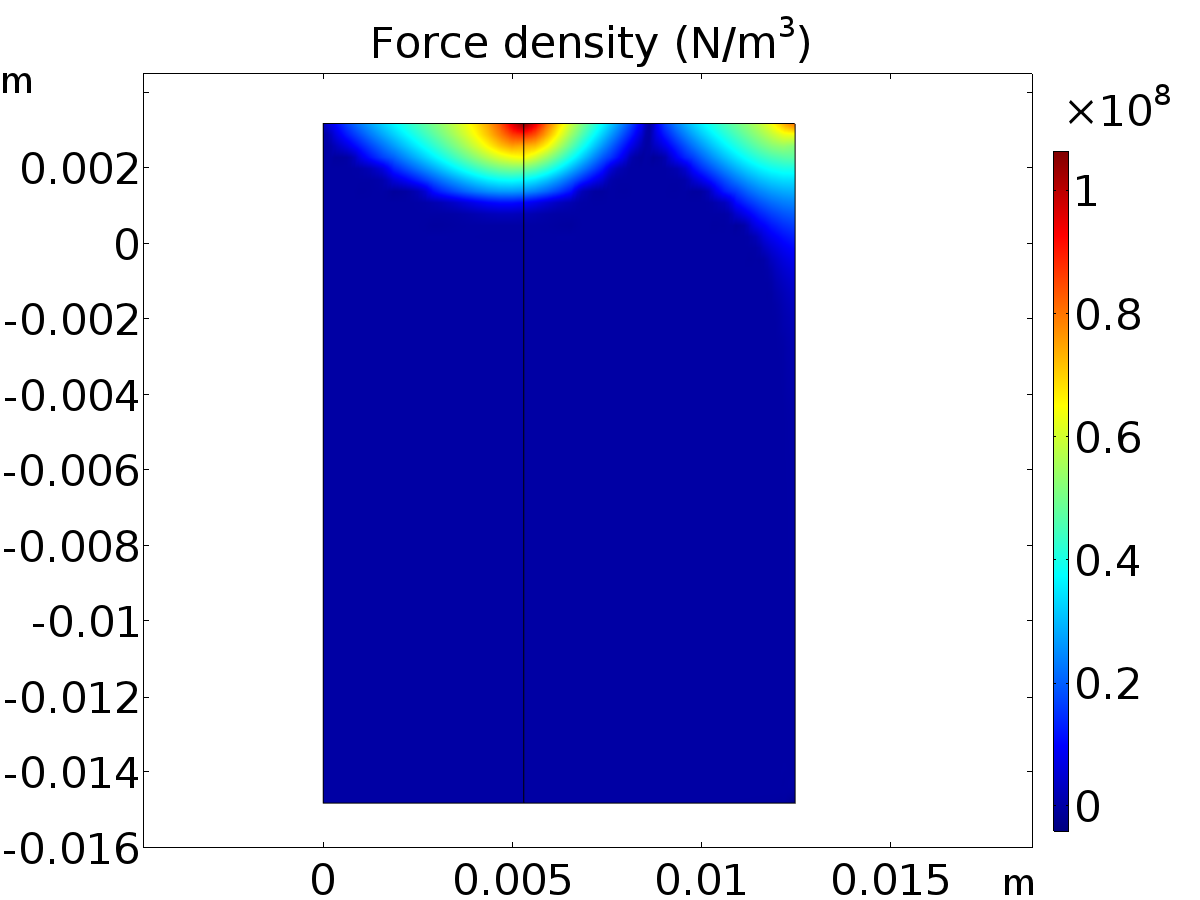}
\includegraphics[width=6 cm]{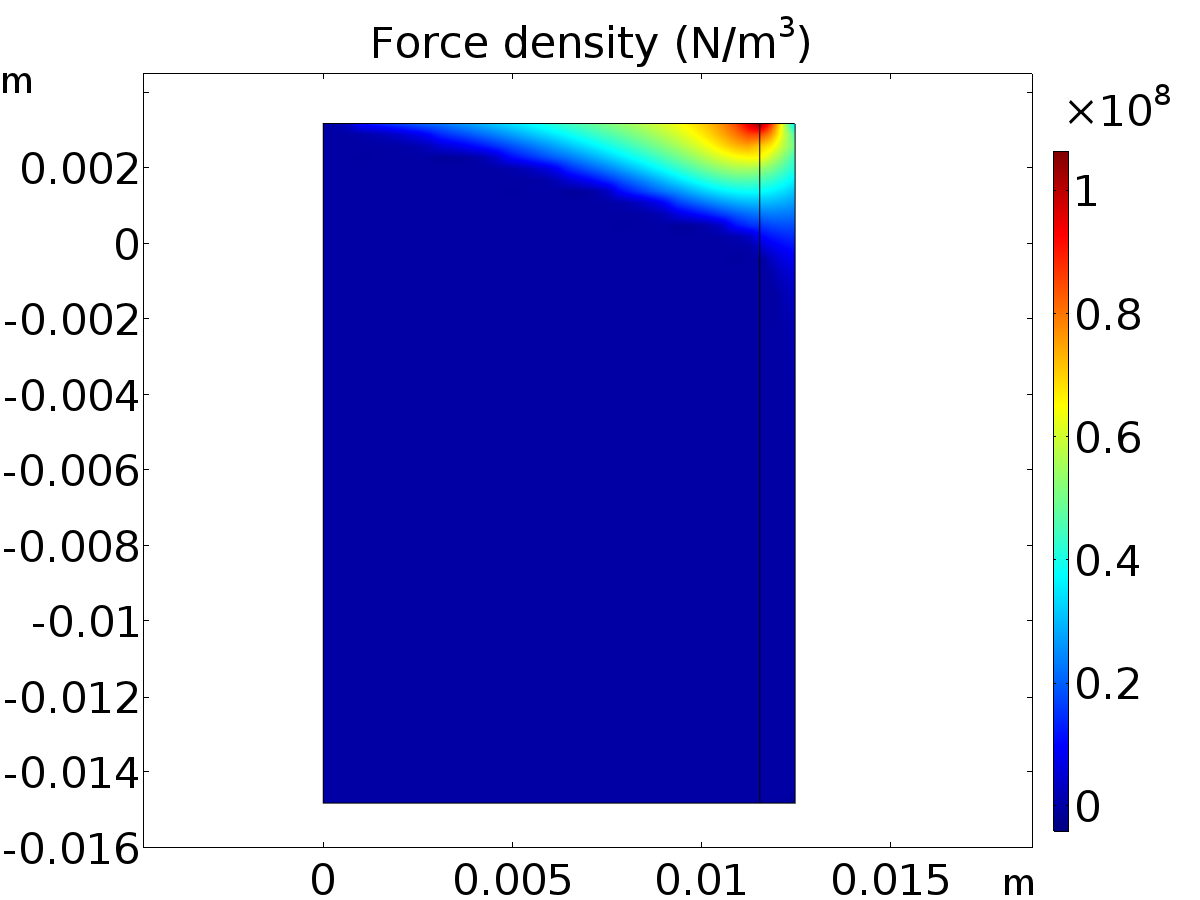}
\includegraphics[width=6 cm]{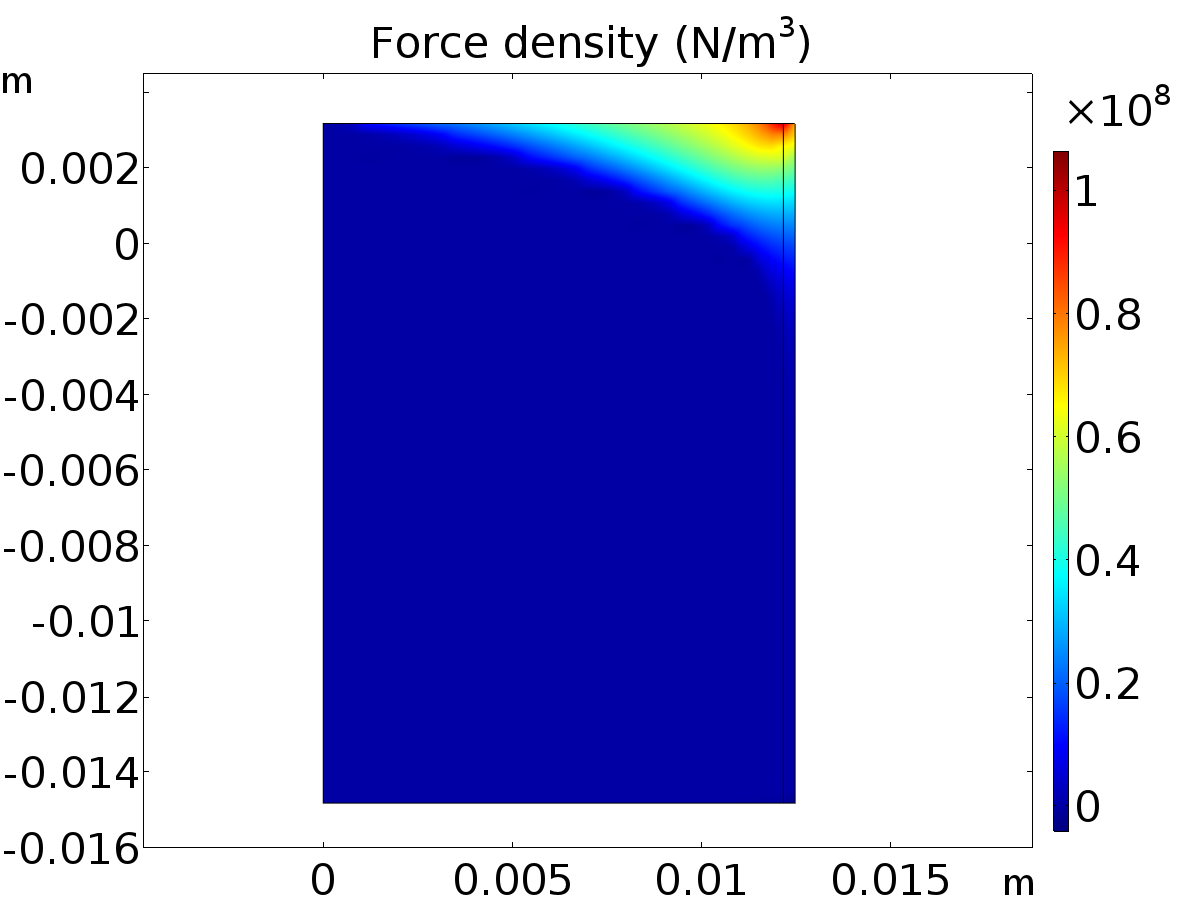}
\caption{\label{fig:maps_magn_up_down}Maps of the force density in the superconductor bulk for the four values of $r_{\rm in}$ corresponding to points a, b, c, d in figure~\ref{fig:magn_up_down}.}
\end{figure}

One can for example check whether using a permanent magnet made of two domains with opposite magnetization has a beneficial effect.
For this purpose, we considered a permanent magnet with the same dimensions as in table~\ref{tab:param}, but we split it into two concentric cylinders of opposite magnetization, the inner one with magnetization directed along $+z$, the outer one with magnetization directed along $-z$. We varied the radius of the inner cylinder $r_{\rm in}$ from 0 to $r_{\rm pm}=\SI{12.5}{\milli\meter}$. The results of the force corresponding to the minimum distance between permanent magnet and superconducting bulk are displayed in figure~\ref{fig:magn_up_down}. 
The two extreme cases $r_{\rm in}=0$ and $r_{\rm in}=\SI{12.5}{\milli\meter}$ correspond to a bulk fully magnetized in the positive and negative $z$ direction: the force is equal to \SI{36.4}{\newton}, as shown in figure~\ref{fig:force}. For intermediate values of $r_{\rm in}$ the forces changes, reaching a maximum for $r_{\rm in}=\SI{11.6}{\milli\meter}$. This can be understood by looking at the maps of the force density in the superconducting bulk, which are shown in figure~\ref{fig:maps_magn_up_down}. The maximum of the force density occurs near the boundary of the two oppositely magnetized permanent magnets. For convenience, in figure \ref{fig:maps_magn_up_down}, the line separating the two oppositely magnetized domains of the permanent magnet is extended to the superconductor domain. On the left of this line, there is is a relatively large region where the force density is still considerable. The total force is maximized when the line is close to the edge of the superconductor, so that the medium-force region (in light blue in the figure) is large, and the high-force region (in red in the figure) is still entirely in the superconductor. If the boundary is situated too close to the edge, part of the high-force region is lost.
The obtained maximum force is about \SI{40}{\newton}, which corresponds to an increase of \SI{10}{\percent} with respect to the case of the permanent magnet magnetized in one direction only. While this increase is to little to justify a practical realization of this configurations, it shows the high practicality of the developed model for investigating different configurations of permanent so as to optimize the levitation system.

\subsubsection{Graded magnetization}
\begin{figure}[h!]
\centering
\includegraphics[width=8 cm]{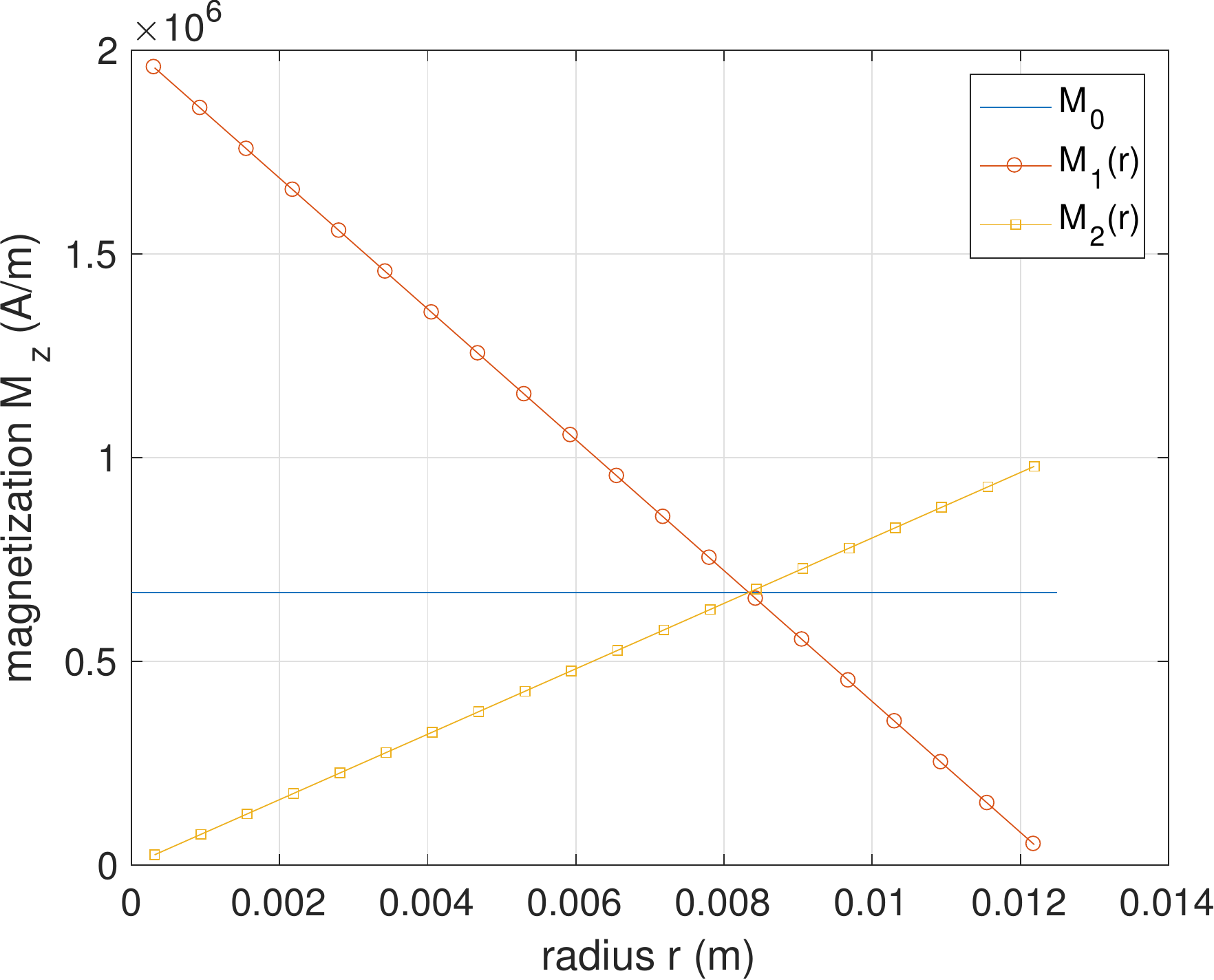}
\caption{\label{fig:Mr}Radial profiles of magnetization $M_z(r)$ used for comparison. The three profiles have the same volume integral in the permanent magnet domain, thus yielding the same magnetic moment.}
\end{figure}

Until now, uniform magnetization in the $z$ direction has been assumed. Can higher levitation forces be obtained by using a vertical magnetization of varying magnitude in the radial direction? In the model, a space-varying magnetization cannot be imposed inside a given domain. However, one can split the permanent bulk cylinder into concentric cylinders, each of them of different magnetization. If the subdivision is sufficiently fine (e.g. 20 subdivisions), this situation approximates well a $z$ magnetization of varying magnitude in the radial direction. In order to make a fair comparison, we considered a variable magnetization $M_z(r)$, such that 
the permanent has the same total magnetic moment as in the case of uniform magnetization. In other words, we assumed that the integral of $M_z(r)$ in the radial direction is equal to that of a constant magnetization $M_0$.
For simplicity, we chose a linear variation $M_z(r)=\alpha r+ \beta$, and we impose that
\begin{equation}
\int\limits_0^{r_{\rm pm}} 2 \pi r M_z(r) {\rm d}r=\int\limits_0^{r_{\rm pm}} 2 \pi r M_0 {\rm d}r=\pi M_0 \frac{{r_{\rm pm}}^2}{2},
\end{equation}
In particular, we consider two situations, where $M_z(r)=0$ at the border ($r=r_{\rm PM}$) and at the center ($r=0$) of the permanent magnet. The two situations are represented in figure~\ref{fig:Mr}, alongside the case of constant magnetization $M_0$. In the figure, the square and circle data points represent the values of magnetization used in each concentric cylinder in the finite-element model.

\begin{figure}[h!]
\centering
\includegraphics[width=6 cm]{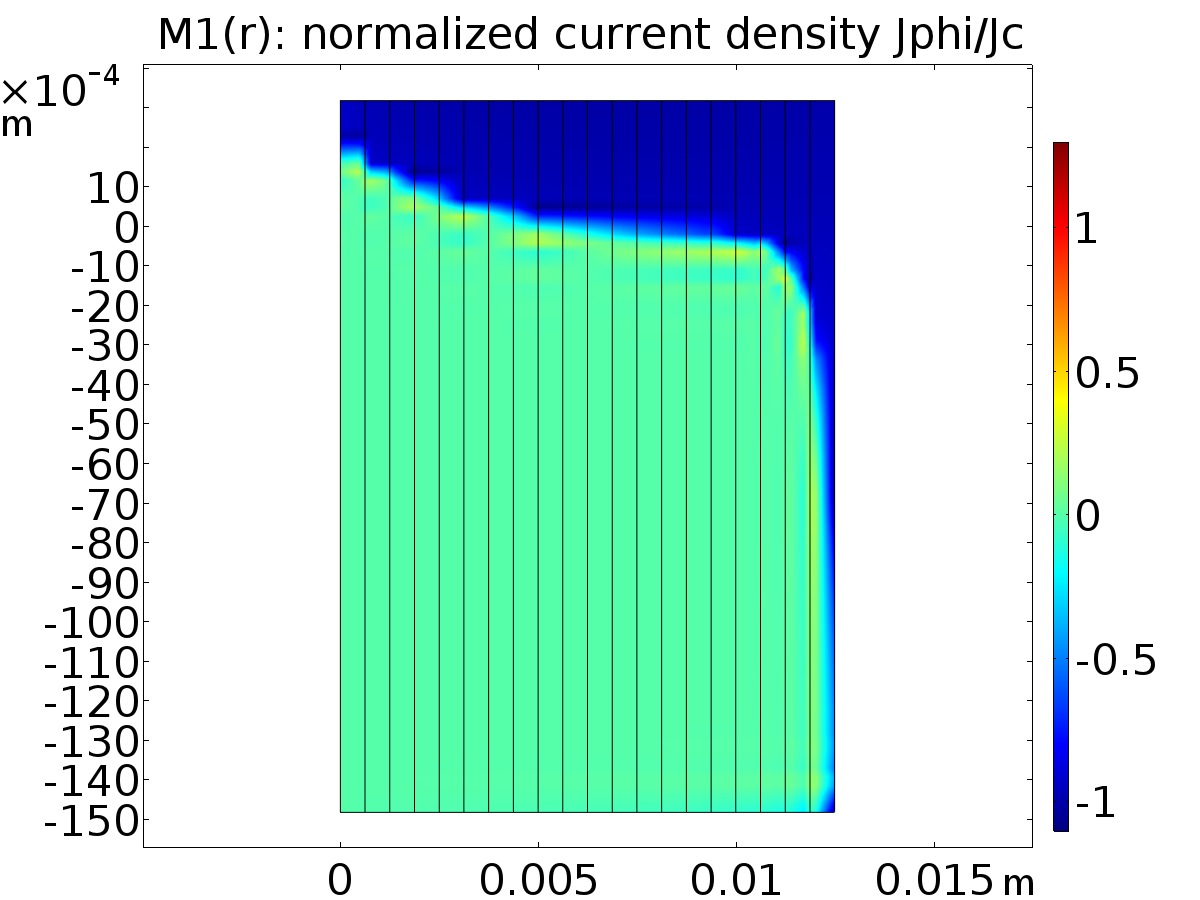}
\includegraphics[width=6 cm]{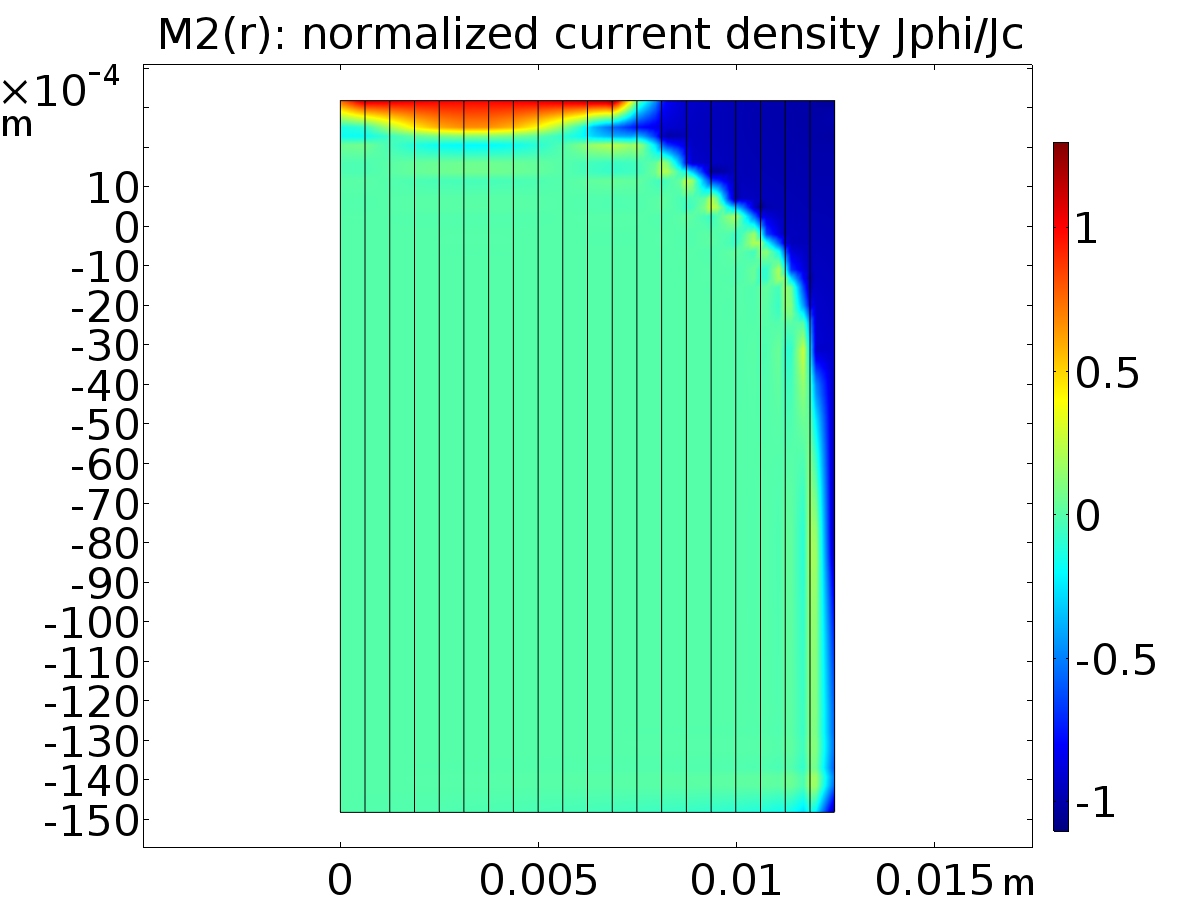}
\caption{\label{fig:Mr_maps}Maps of the normalized current density obtained with the radial profiles of magnetization $M_1(r)$ and $M_2(r)$ shown in figure~\ref{fig:Mr}. The distributions are taken at the instant of minimum distance (\SI{0.1}{\milli\meter} between the superconductor and the permanent magnet.}
\end{figure}

\begin{figure}[h!]
\centering
\includegraphics[width=6 cm]{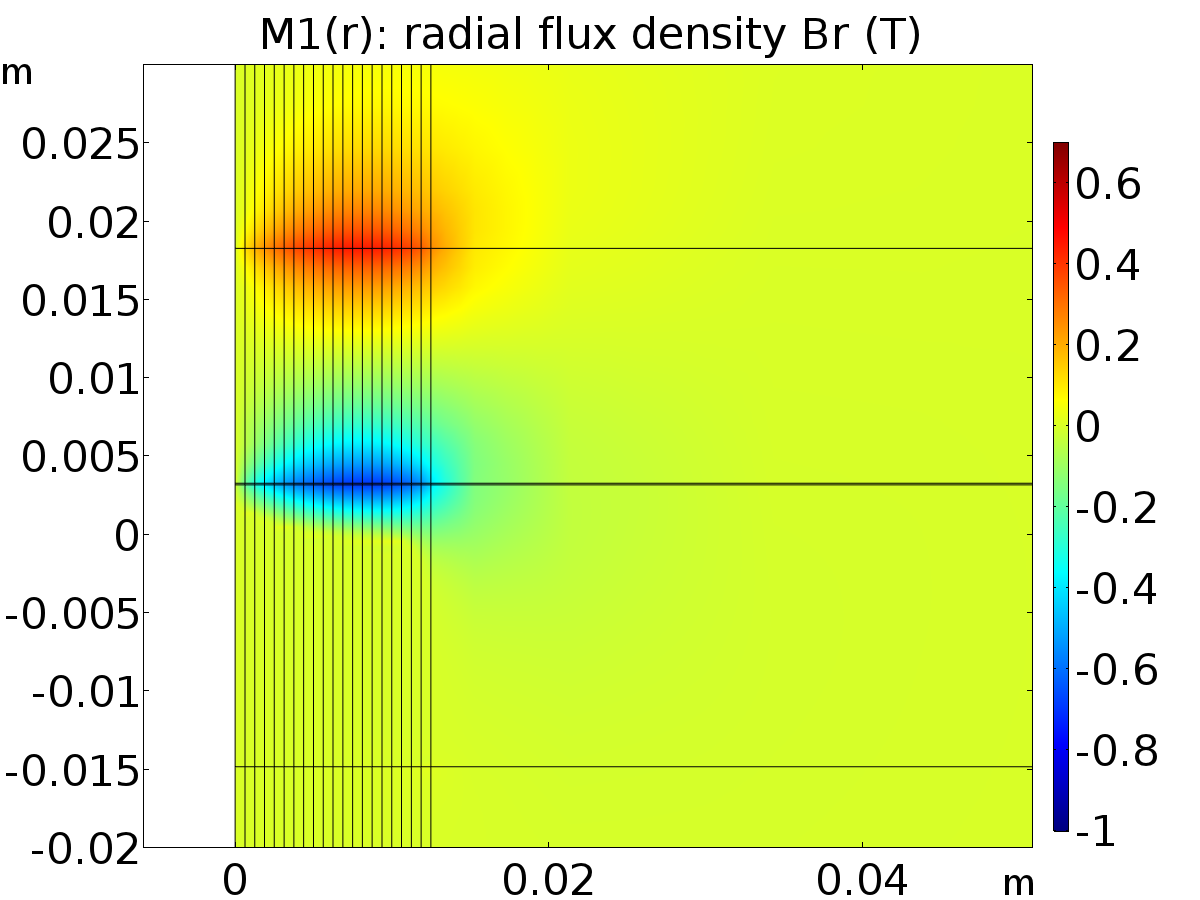}
\includegraphics[width=6 cm]{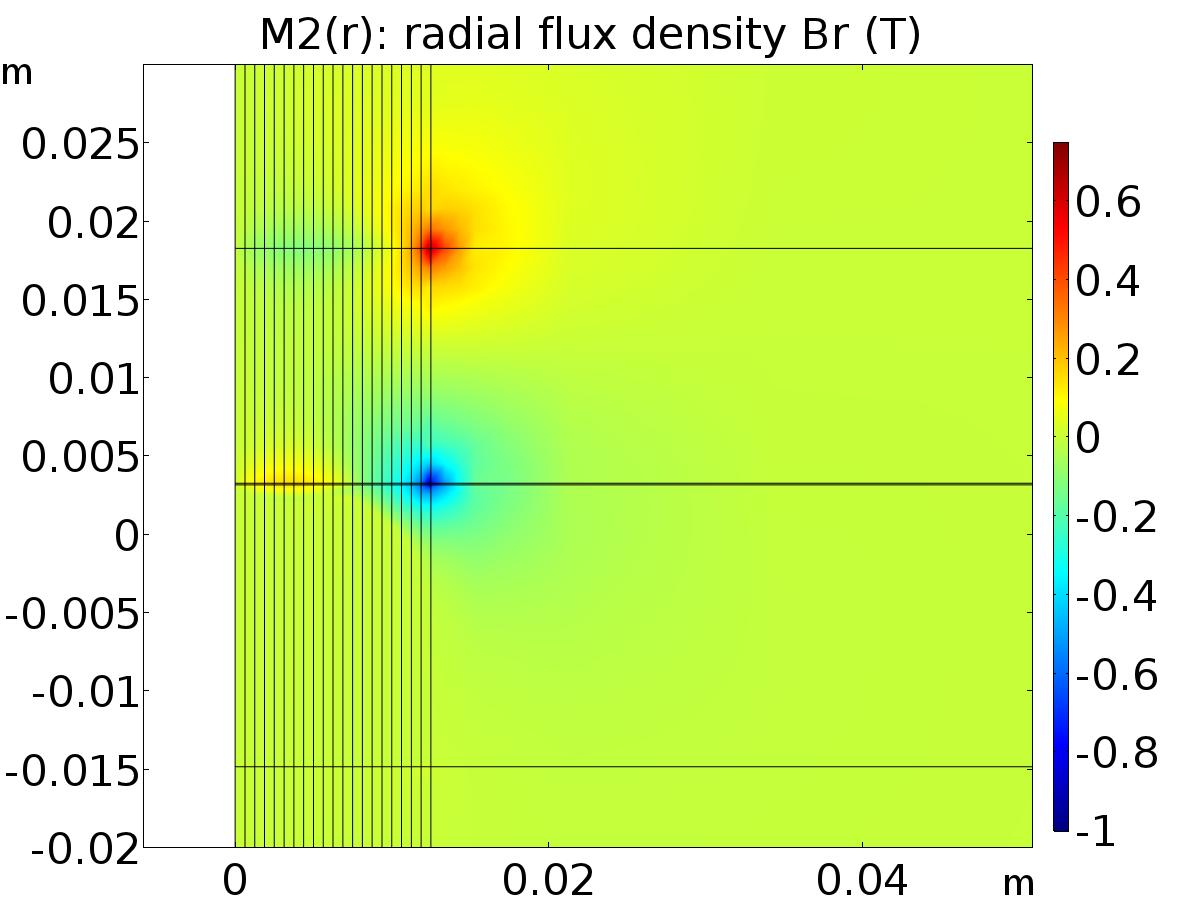}
\caption{\label{fig:Bradial_maps}Maps of the radial component of the magnetic flux density obtained with the radial profiles of magnetization $M_1(r)$ and $M_2(r)$ shown in figure~\ref{fig:Mr}. The distributions are taken at the instant of minimum distance (\SI{0.1}{\milli\meter} between the superconductor and the permanent magnet.}
\end{figure}

The magnetization $M_1(r)$ gives a maxim force of \SI{70}{\newton}, which is almost twice as high as that obtained with a constant magnetization $M_0$. On the contrary, the magnetization $M_2(r)$ gives a magnetization of only \SI{26}{\newton}. The different results can be understood by looking at the corresponding current density and radial field distributions (taken at the instant of minimum distance (\SI{0.1}{\milli\meter}) between superconductor and permanent magnet), which are displayed in figure~\ref{fig:Mr_maps} and~\ref{fig:Bradial_maps}, respectively. The magnetization $M_1(r)$ has a very large magnetization near the $z$-axis, and the densely packed flux lines of this magnetization are compressed between the superconductor bulk and the permanent magnet. This results in  large radial field component and a large induced current area in the superconductor, which in turn results in a large levitation force. On the contrary, with the magnetization $M_2(r)$ profile, most most of the magnetization is situated at the periphery of the permanent magnet: when the permanent magnet approaches the superconductor bulk, most of the flux lines easily escape on the side ($r > r_{\rm PM}$), and there is less electric field that can induce current and hence levitation force.
 
\newpage
\subsubsection{Free-fall oscillations}\label{subsec:grav}
\begin{figure}[h!]
\centering
\includegraphics[width=8 cm]{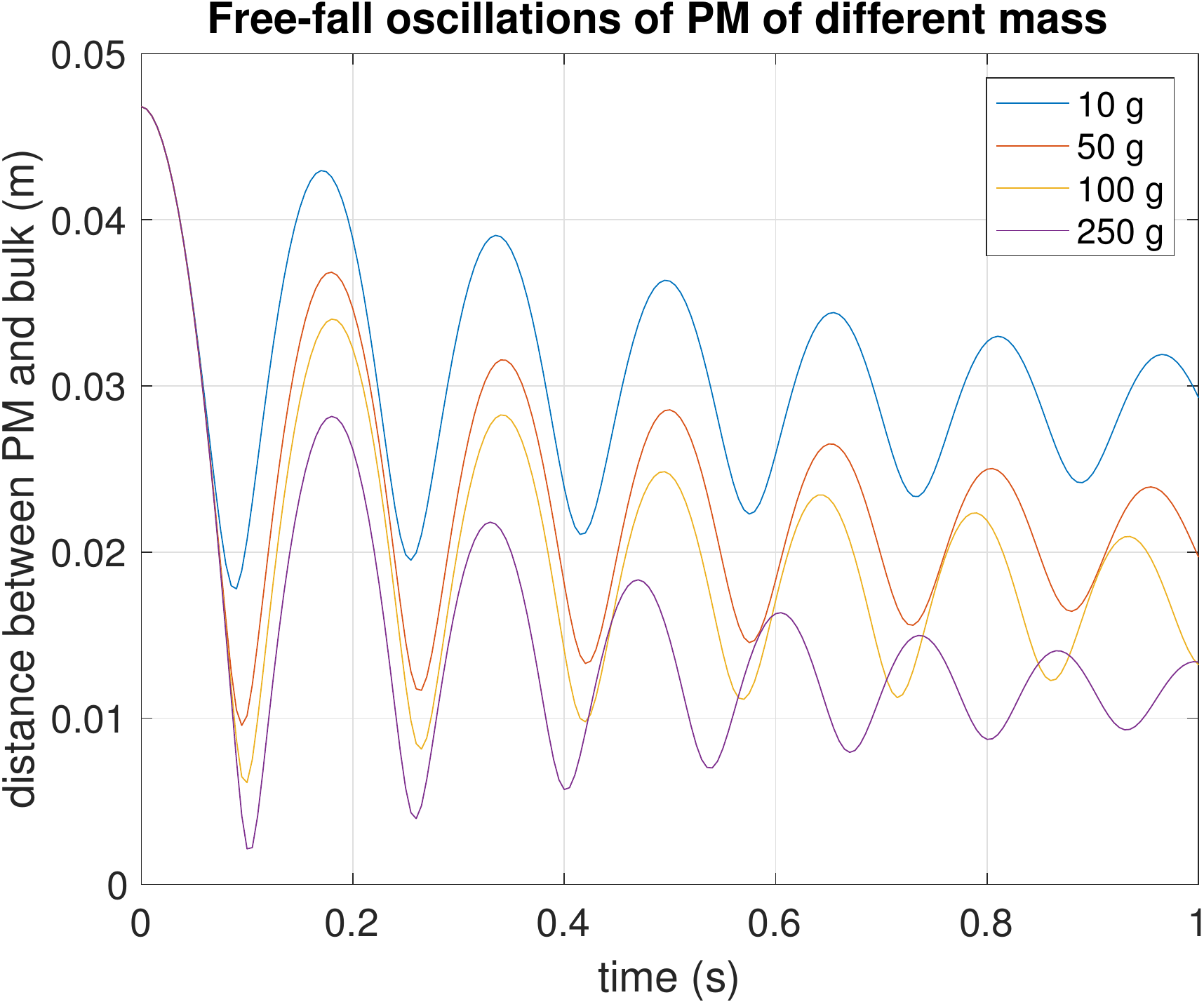}
\caption{\label{fig:grav}Free-fall oscillations of permanent magnets of different mass.}
\end{figure}

Finally, we considered the case of free-fall oscillations. Instead of imposing a given displacement to the permanent magnet, we let it fall, starting with zero initial velocity,  from a distance $z_0=\SI{46.81}{\milli\meter}$ from the top of the superconductor bulk, using the approach described in section~\ref{subsec:grav}. The drag force can be easily inserted in the model, but it much smaller compared to the electromagnetic breaking effect produced by the superconductor, and can be neglected (as also reported in~\cite{Zhou:PhysC06}). The permanent magnet oscillates above the superconductor bulk, as shown in figure~\ref{fig:grav}. The amplitude and frequency of those oscillations depend on the mass of the permanent magnet, with the frequency being proportional to the square root of the mass, as in a harmonic oscillator.

\section{Conclusion}
In this paper, we presented a new finite-element method approach for modeling levitation phenomena with superconducting bulks and permanent magnets, based on the coupling of the $H$-magnetic field and the Arbitrary Lagrangian-Eulerian formulations. This approach allows modeling the electrodynamic behavior of the system with the actual movement of the permanent magnet domain, without the need of setting time-dependent boundary conditions for simulating such movement in a fixed-geometry problem, as typically done by other models presented in the literature. The model has been implemented in Comsol Multiphysics, with the MFH and ALE modules. This allows for much greater flexibility and easiness of use of the model.

The model has first been validated by comparing its results for a 2D axisymmetric problem with those of a recently proposed model based on the $A-\phi$ formulation of the eddy current problem, which has been in turn validated against experimental data, both in the case of field-cooled and zero field-cooled levitation.
Then the model has been used to simulate situations of interest for practical applications, for example permanent magnet with opposite magnetizations of different size, graded magnetization and free-fall oscillations.
The obtained results of levitation forces have been interpreted with the help of magnetic field and current density distributions and demonstrated the effectiveness and the potential of the model as a primary tool for studying levitation phenomena in practical applications.

\end{document}